\documentclass[12pt,a4paper,dvips]{article}
\usepackage{a4p}
\usepackage{cite,mcite}
\usepackage{graphicx}
\usepackage{physics}
\usepackage{l3_title,ifthen}
\usepackage{Lep}
\date{September 29, 1999}
\preprint{99-137}
\newlength{\capindent}
\newlength{\capwidth}
\newlength{\figwidth}
\newcommand{\icaption}[2][!*!,!]{\hspace*{\capindent}%
  \begin{minipage}{\capwidth}
    \ifthenelse{\equal{#1}{!*!,!}}%
      {\caption{#2}}%
      {\caption[#1]{#2}}
  \end{minipage}}
\def\gappeq{\mathrel{\rlap {\raise.4ex\hbox{$>$}}
{\lower.6ex\hbox{$\sim$}}}}

\def\lappeq{\mathrel{\rlap{\raise.4ex\hbox{$<$}}
{\lower.6ex\hbox{$\sim$}}}}

\newcommand {\Be}{\begin{equation}}   
\newcommand {\Ee}{\end{equation}}     
\newcommand {\eqref}[1]{equation~(\ref{#1})}

\newcommand {\Figref}[1]{Figure~\ref{fig:#1}}
\newcommand {\Tabref}[1]{Table~\ref{tab:#1}}

\def\chna   {\mathrm{\tilde{\chi}_1^0}}
\def\qst    {\mathrm{\tilde{t}_1}}
\def\qsb    {\mathrm{\tilde{b}_1}}
\def\qstr   {\mathrm{\tilde{t}_R}}
\def\qsbr   {\mathrm{\tilde{b}_R}}
\def\qast   {\mathrm{\bar{\tilde{t}}_1}}
\def\qasb   {\mathrm{\bar{\tilde{b}}_1}}
\def\snu    {\mathrm{\tilde{\nu}}}

\newcommand {\evis}{$E_{\mathrm{vis}}$}
\newcommand {\dbt}{$D_{\mathrm{Btag}}$}
\newcommand {\ettj} {$E_\mathrm{TTJ}$}

\begin{document}
\begin{titlepage}
\title{\boldmath
Searches for Scalar Quarks in $\mathrm{e^+e^-}$ Interactions \\
at  $\sqrt{s}$ = 189 \gev{}
}
\author{The L3 Collaboration}

\vspace*{-10mm}

\begin{abstract}
Searches for scalar top and scalar bottom quarks, as well 
as for mass-degenerate scalar quarks of the first two 
families are performed at 189 \gev{} centre-of-mass energy
with the L3 detector at LEP. No signals are observed.
Model-independent limits on the scalar top production cross sections are
determined  for the decay modes $\qst \to \mathrm{c} \chna$
and $\qst \to \mathrm{b} \ell \snu$.
For scalar quarks of the other flavours
$\tilde{\mathrm{q}} \to \mathrm{q} \chna$ decays are
considered. Within the framework of the Minimal Supersymmetric Standard
Model mass  limits are set at 95\% C.L.
for these particles. Indirect limits on the gluino mass
are also derived.
\end{abstract}
%
%
\vspace*{-10mm}
\submitted

\end{titlepage}

\clearpage
%
%
\section{Introduction}
In the Minimal Supersymmetric extension of the Standard Model
(MSSM)~\cite{susy}
for each helicity state of Standard Model (SM) quarks, $\mathrm{q_{L,R}}$,
there is a
corresponding scalar SUSY partner $\mathrm{\tilde{q}_{L,R}}$. 
Generally, the left, $\mathrm{\tilde{q}_{L}}$,
and right, $\mathrm{\tilde{q}_{R}}$, eigenstates mix to
form mass eigenstates. The mixing is
proportional to the corresponding SM quark mass and to the
parameter $a_{\mathrm{q}}=A_{\mathrm{q}}-\mu\cot\beta$ for up
type quarks and $a_{\mathrm{q}}=A_{\mathrm{q}}-\mu\tan\beta$
for down type ones. 
$A_{\mathrm{q}}$ is the trilinear coupling
among scalars, $\mu$ the Higgsino mass parameter and 
$\tan\beta$  the ratio of the vacuum expectation values
of the two Higgs fields. For
the first two generations of
scalar quarks (squarks) the weak eigenstates are also mass eigenstates
to a good approximation. However, this does not
hold for the squarks of the third family.
Due to the heavy top quark the 
$\mathrm{\tilde{t}_{L}} - \mathrm{\tilde{t}_{R}}$ mixing
is enhanced,
leading to a large  splitting between the two mass eigenstates.
This is usually expressed in terms of 
the  mixing angle, $\theta_{\mathrm{LR}}$.
The lighter scalar top (stop) quark
\begin{equation}
\mathrm{\tilde{t}_1} = \mathrm{\tilde{t}_L} \cos\theta_{\mathrm{LR}} +  
\mathrm{\tilde{t}_R} \sin\theta_{\mathrm{LR}}
\end{equation}
can thus be well within the discovery range of LEP.
Furthermore, if $\tan\beta \gappeq$ 10,
large $\mathrm{\tilde{b}_L - \tilde{b}_R}$ mixing
occurs. This may lead to a scalar bottom (sbottom) quark, $\qsb$, 
also light enough to be accessible at LEP.

In the present analysis, R-parity conservation
is assumed, which implies that SUSY particles (sparticles) are 
produced in pairs; heavier sparticles decay into
lighter ones and the Lightest Supersymmetric 
Particle (LSP) is stable.
In the MSSM the best
LSP candidate is the weakly interacting lightest 
neutralino, $\chna$.

Squark pair production at LEP proceeds via the exchange of 
Z/$\gamma$ bosons in the  $s$-channel. The production cross section is 
governed by two free parameters: the
squark mass and the mixing angle,
$\theta_{\mathrm{LR}}$~\cite{wien}.
At $\cos\theta_{\mathrm{LR}}\sim$ 0.57  the
stop decouples from the $\mathrm{Z}$ and the cross
section  is  minimal. The corresponding value is
$\cos\theta_{\mathrm{LR}}\sim$ 0.39
for the sbottom.
The cross section reaches the maximum at  
$\cos\theta_{\mathrm{LR}}$=1 when the light
squark mass eigenstate is the weak eigenstate.

At LEP energies the most important stop decay
channels are: $\qst \to \mathrm{c} \chna$, 
$\mathrm{b} \nu_{\ell}\tilde{\ell}$, $\mathrm{b}\ell\tilde{\nu_{\ell}}$, and
$\mathrm{b}\tilde{\chi}^{+}_1$,
where the $\tilde{\ell}$ and  $\tilde{\nu_{\ell}}$
are the supersymmetric
partners of the charged leptons and neutrinos, and
the $\chna$ and $\mathrm{\tilde{\chi}^{+}_1}$ are the lightest
 neutralino and chargino, respectively.
The $\qst \to \mathrm{b} \tilde{\chi}^{\pm}_1$ decay
channel is the dominant one when kinematically allowed.
However, the current limits on the chargino mass~\cite{chargino}
preclude this decay to occur, except for a small region in the MSSM
parameter space with the common scalar mass ($m_0$)
from 60 to 90 \gev{}.
Similarly, the $\qst \to \mathrm{b} \nu_{\ell}\tilde{\ell}$ decay
is precluded by the current limits \cite{slepton} on 
charged scalar lepton masses.
The stop analysis is performed considering
the $\qst \to \mathrm{c} \chna$ and
$\qst \to \mathrm{b}\ell\tilde{\nu_{\ell}}$ 
decay channels, with $\tilde{\nu_{\ell}}$ decaying invisibly
$\tilde{\nu_{\ell}} \to \nu_{\ell} \chna$.
Since the $\qst \to \mathrm{c} \chna$ 
is a flavour changing weak decay, the $\qst \to
\mathrm{b}\ell\tilde{\nu_{\ell}}$
channel is dominant when kinematically allowed. Therefore
the two decay modes are investigated each with
the assumption of 100\% branching fraction.
For the stop three-body decay channel $\qst \to
\mathrm{b}\ell\tilde{\nu_{\ell}}$, two scenarios are
considered:
$\ell$ being e, $\mu$ or $\tau$ with equal probabilities or pure
$\tau$.
The latter occurs at high  $\tan\beta$
values. 

For sbottom, as well as for the first two generations of squarks,
the $\tilde{\mathrm{q}} \to \mathrm{q} \chna$ decay mode is
investigated  under the assumption of
100\% branching fraction.

Since the stop two-body decay $\qst \to \mathrm{c} \chna$ is a second
order weak decay, the lifetime of the $\qst$  is larger than the
typical
hadronisation time of $10^{-23}$s. The $\qst \to \mathrm{b} \ell \snu$
decay proceeds via a
virtual chargino exchange and the lifetime is also
expected to be larger than the hadronisation time.
Thus the stop will first hadronise and then decay. 
For the sbottom the situation depends on the gaugino-higgsino
content of  the neutralino:  for a gaugino-like neutralino
the sbottom lifetime is expected to be larger than 
the hadronisation time. In the present analysis a
`hadronisation before decay' scenario is followed. 
Although hadronisation does not change the final event
topology, it affects the track multiplicity, 
the jet properties and the event shape.

The present study supersedes previous L3
limits on stop and sbottom quark  productions\cite{L3stop}.
Searches for supersymmetric quarks have been performed by other
LEP~\cite{limits} and by TEVATRON~\cite{CDF,D0} experiments.

%
\section{Data Samples and Simulation }

The data used in the present analysis were collected in 
1998 at $\sqrt{s}$=189 \gev{} using the L3 detector~\cite{l3}.
The total integrated luminosity is 176.4~pb$^{-1}$.

Monte Carlo (MC) samples of squark events  are generated using a
PYTHIA based event generator~\cite{sqgen}. 
The 
squark mass has been varied from 45 \gev{} up to the kinematical limit
and the $\chna$ mass from 1 \gev{} to
$M_{\qst} - $2 \gev{}
or to $M_{\qsb} - $5 \gev{}
for the stop and sbottom two-body decay modes.
The $\qst \to \mathrm{b} \ell \snu$ and 
$\qst \to \mathrm{b} \tau \snu$ channels are generated with
$\snu$ mass from 43 \gev{} to $M_{\qst} - 7$ \gev{}.
In total 160 samples are generated, each with at least
2000 events.

The following MC programs
are used to estimate the Standard Model backgrounds:
PYTHIA ~\cite{pythia} for $\mathrm{e^+e^- \to q\bar{q}}$, 
 $\mathrm{e^+e^- \to ZZ}$ and
$\mathrm{e^+e^- \to Z e^+e^-}$,
KORALZ~\cite{koralz} for $\mathrm{e^+e^- \to \tau^+ \tau^-}$,
KORALW~\cite{koralw} for $\mathrm{e^+e^- \to W^+W^-}$,
EXCALIBUR~\cite{excali} for $\mathrm{e^+e^- \to
W^{\pm}e^{\mp}\nu}$,
PHOJET~\cite{phojet} for $\mathrm{e^+e^- \to e^+e^- q \bar{q}}$ and
DIAG36~\cite{diag36} for $\mathrm{e^+e^- \to e^+e^-  \tau^+  \tau^-}$.
The number of simulated events for each background
process exceeds by 100 times the statistics of the collected
data samples except for the two-photon collision
processes, for which the MC statistics amounts to only 
twice the data.

The response of the L3 detector is simulated using the GEANT 3.15 
package~\cite{geant}. It takes into account effects of
energy loss, multiple scattering and showering in the detector 
materials and in the beam pipe. Hadronic interactions  are simulated
with the GHEISHA program ~\cite{geisha}.

%
%

\section{Event Preselection}
The signal events of $\qst \to \mathrm{c} \chna$ and
$\qsb \to \mathrm{b} \chna$ contain
two high multiplicity
acoplanar jets originated from c or b-quarks.
In addition, two charged leptons are present in the
$\qst \to \mathrm{b} \ell \snu$ decay channel. 
The neutralinos and sneutrinos in the final state
escape detection leading to missing energy in the event.
A common preselection is applied to obtain a sample of unbalanced
hadronic events.
The events have to fulfil the following requirements:
more than four tracks;
at least 10 but not more than 40 calorimetric clusters;
a visible energy, \evis, between 5 \gev{} and  150 \gev{};
an energy deposition in the forward calorimeters
less than 10 \gev{} and a total energy in the $30^{\circ}$ 
cone around the beam pipe less
than 0.25$\times$\evis;
a transverse missing momentum, $P{\mathrm{^{miss}_T}}$,
 greater than 2 \gev{} and a sinus of the
polar angle of the missing momentum,
$\sin\theta_{\mathrm{miss}}$, greater than
0.2.

\begin{figure}[h]
\vspace*{-40mm}
\hspace*{-5mm}
\includegraphics[bb=3 3 525 525,width=170mm,height=185mm,
    clip=true,draft=false]{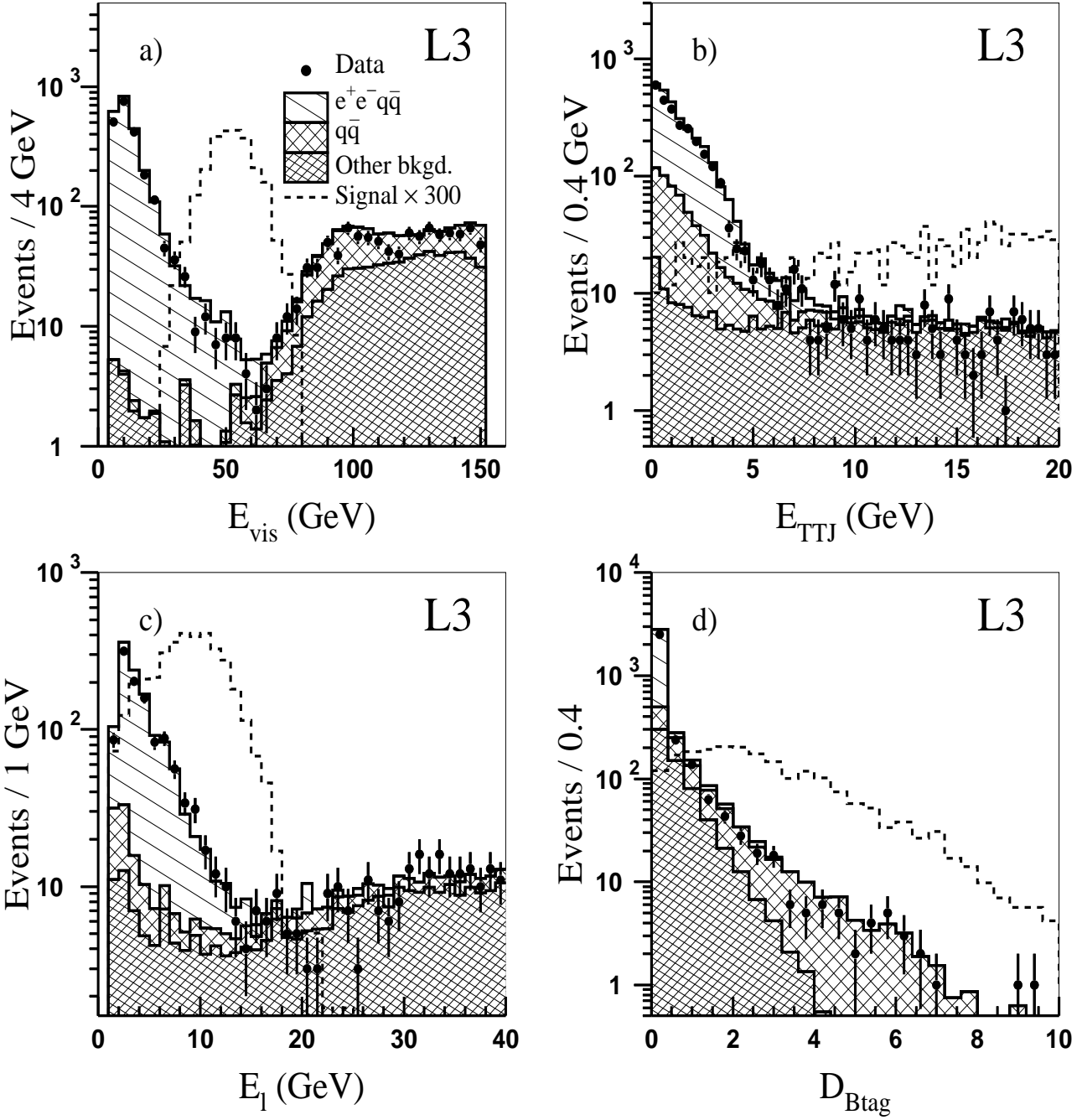}
\caption{\label{fig:kin_189} 
Distributions of a) \evis, b) \ettj \ (see text),
c) the most energetic lepton energy $E_{\ell}$, and
d) b-tagging event discriminant \dbt \
for data and  MC events after
preselection.
Contributions from $\mathrm{e^+e^- q\bar{q}}$,
$\mathrm{q\bar{q}}$ and other backgrounds, dominated
by $\mathrm{W^+W^-}$ production, are given separately.
The distributions for expected signal events of
$\qstr \to \mathrm{c} \chna$ with 
$M_{\qstr}$=90 \gev{},
$M_{\chna}$=60 \gev{} (a,b),
$\qstr \to \mathrm{b} \ell \snu$ \ with
$M_{\qstr}$=90 \gev{},
$M_{\snu}$=70 \gev{} (c) and
$\qsbr \to \mathrm{b} \chna$ with
$M_{\qstr}$=90 \gev{},
$M_{\chna}$=60 \gev{} (d) 
are also shown.}
\end{figure}

After the preselection 3110 events are retained,
compared with 3514~$\pm$~48 
expected from the SM processes, which are dominated by 
two-photon interactions.
\Figref{kin_189} shows the distributions of 
\evis; the absolute value of the 
total momentum of the two jets projected
onto the direction perpendicular to the
thrust axis computed in the transverse plane, \ettj;
the energy of the most energetic lepton, 
$E_{\mathrm{\ell}}$
and the b-tagging event discriminant, \dbt.
\dbt \
is defined as the negative log-likelihood of the   
probability for the event to be consistent with light quark
production~\cite{btag}.
After preselection the data and MC  are in 
good agreement. The discrepancy in the total 
number of data and MC events is localised
in the low \evis \ region, which is dominated by two-photon
processes. This effect is taken into account by assigning
a systematic error of 10--20\% on the two-photon cross section.

\section{Selection Optimisation}
The kinematics of the signal events strongly
depend on the mass difference
between squark and neutralino, $\Delta M$=$M_{\mathrm{\tilde{q}}} -
M_{\chna}$.  In the very low $\Delta M$ region,
the visible
energy and track multiplicity are low. Therefore, signal events are
difficult to separate from the two-photon interactions. For high
$\Delta M$ values,
signal events will be similar to $\mathrm{W^+W^-}$,
$\mathrm{W^{\pm}e^{\mp}\nu}$ or
$\mathrm{ZZ}$ final
states. The most favourable
region for the signal and background separation is expected at 
$\Delta M$=20--40 \gev{}. 

To cope with the various background sources, the searches 
are performed independently in different $\Delta M$ regions.
For $\qst \to \mathrm{c} \chna$ and $\qsb \to \mathrm{b} \chna$
decays  four selections have been optimised. These
selections typically cover $\Delta M$ regions of: 5--10 \gev{},
10--20 \gev{}, 20--40 \gev{} and $\gappeq$ 40 \gev{}.
In case of 
$\qst \to \mathrm{b} \ell \snu$  decays three selections are devised 
for each lepton flavour. These selections cover the
$\Delta M$ = $M_{\mathrm{\tilde{q}}} - M_{\snu}$
region consistent with the limit $M_{\snu} \gappeq$43 \gev{}
from LEP1~\cite{LEP1snu}.

The following kinematic variables are used in the selections:
Lower cuts on
\evis, 
$P{\mathrm{^{miss}_T}}$ and $P{\mathrm{^{miss}_T}}$/\evis~
separate signal  
from two-photon background, whereas an upper cut 
on \evis \ removes 
$\mathrm{W^+W^-}$,  $\mathrm{W^{\pm}e^{\mp}\nu}$,
$\mathrm{ZZ}$ and $\mathrm{Ze^+e^-}$ events. A cut on
$\sin\theta_{\mathrm{miss}}$
rejects $\mathrm{e^+e^-q\bar{q}}$ events. 
Cuts on jet acollinearity and acoplanarity
reduce the $\mathrm{q\bar{q}}$ contribution.
A veto on the energy deposition in the 50$^{\circ}$ azimuthal sector
around
the missing momentum direction suppresses $\tau^+\tau^-$
and $\mathrm{q\bar{q}}$
events.
The $\mathrm{W^+W^-}$ production, where one $\mathrm{W}$
decays leptonically and $\mathrm{W^{\pm}e^{\mp}\nu}$ events
are removed by vetoing energetic isolated leptons.
The cut on \ettj \ suppresses
$\mathrm{e^+e^-q\bar{q}}$, $\mathrm{q\bar{q}}$ as well as
$\mathrm{W^+W^-}$ backgrounds.

For the selections of $\qsb \to \mathrm{b} \chna$ and
$\qst \to \mathrm{b} \ell \snu$  signal events,
cuts are applied on
the event b-tagging variable \dbt.

At least one isolated lepton is required
in the case of $\qst \to \mathrm{b} \ell \snu$  decays.
An electron is 
isolated if the calorimetric energy deposition in a $10^{\circ}$ 
cone  around its direction
is less than 2 \gev{}. Muon isolation requirement implies
an energy deposition in the cone between $5^{\circ}$ to $10^{\circ}$
around its direction of less than 2 \gev{}. A tau is  
isolated when the calorimetric energy deposition in the cone
between $10^{\circ}$ to $20^{\circ}$ around its direction is less than
2 \gev{}
and less than 50\% of the tau energy. Furthermore, the
energy deposition in a cone between $20^{\circ}$ to $30^{\circ}$
must be less than 60\% of the tau energy. 
Finally, a lower cut on the energy of the most
energetic lepton in the event is applied in order to suppress mainly the
two-photon and  the $\mathrm{q\bar{q}}$ backgrounds.

The cut values on the kinematic variables
are chosen by an optimisation procedure for the
different $\Delta M$ regions.
The procedure  minimises the average
limit for an infinite number of
trials assuming only background contributions
~\cite{sens}.
For each signal mass point, the optimal selection or combination
of selections is chosen.

The expected signal efficiencies for a 90 \gev{} stop 
and sbottom
at various $\Delta M$ values
are given in  \Tabref{effic} together with the SM background
expectations.

\section{Systematic Errors}

The errors arising from the signal MC statistics 
vary from 3\% to 8\% for the stop and  from 3\% to 7\% for the 
sbottom depending on selection efficiencies.

The main systematic errors on the signal selection efficiency
arise from the uncertainties in the squark
production, hadronisation and decay scheme. 
We have studied  the following sources of systematic errors:

\begin{itemize}

\item 
The squark signals are generated assuming 
$\cos\theta_{\mathrm{LR}}$=1. However, as their coupling to the
$\mathrm{Z}$ depends on $\cos\theta_{\mathrm{LR}}$, the
initial
state radiation spectrum is also 
mixing angle dependent. The maximal influence of this source 
has been evaluated by generating signal samples
with the values of $\cos\theta_{\mathrm{LR}}$ when the squarks
decouple from the Z. 
The largest decrease in the selection 
efficiencies, 4\% for  stop and 6\% for sbottom, is 
observed at low $\Delta M$ $\sim$ 5--10 \gev{}. With increasing
 $\Delta M$ the selection efficiencies are less affected
by this source of systematics. At $\Delta M$ $\sim$ 70 \gev{}
the error is estimated to be negligible. 
Conservatively,
for the limit calculation we use the  efficiencies
obtained at decoupling values of $\cos\theta_{\mathrm{LR}}$.

\item The invariant mass available
for spectator quarks has been assumed to be 
$M_{\mathrm{eff}}$=0.5 \gev{}~\cite{fermim}.
The  hadronic energy and track multiplicity of the event
depend on the value of this variable. A
variation of $M_{\mathrm{eff}}$ from 0.25 \gev{} to
0.75 \gev{}~\cite{fermim}
results in $4-12\%$ relative change in efficiency for stop
and $6-8\%$ for sbottom.

\item 
For the hadron containing a squark, the Peterson fragmentation
scheme~\cite{Peterson} is used with  the parameter
$\epsilon_{\mathrm{\tilde{q}}}$
propagated from $\epsilon_{\mathrm{b}}$ such that
$\epsilon_{\mathrm{\tilde{q}}}=
\epsilon_{\mathrm{b}}m_\mathrm{b}^2/m_{\mathrm{\tilde{q}}}^2$ with
$\epsilon_{\mathrm{b}}=0.0035$~\cite{epsilon} and 
$m_\mathrm{b}$=5 \gev{}.
The $\epsilon_{\mathrm{b}}$ is varied in the range from  0.002 to 
0.006~\cite{epsilon}.
This induces $5-12\%$ and $2-6\%$ changes in the selection
efficiencies for $\qst$  and $\qsb$, respectively.

\item For the $\qst \to \mathrm{c} \chna$ decays the
uncertainty on the c-quark fragmentation parameter $\epsilon_\mathrm{c}$
results in a $1-4\%$ change in efficiency when
$\epsilon_\mathrm{c}$ is varied from 0.02 to 0.06~\cite{epsilon}.  
The central value is chosen to be
$\epsilon_\mathrm{c}$ = 0.03~\cite{epsilon}.

\item For the stop three-body decay mode 
$\qst \to \mathrm{b} \ell \snu$,  the weak structure of
the decay matrix element \cite{matrixel} is taken into account.
The related possible source of systematics has
been evaluated by generating signal events
with only a phase-space model. The selection efficiencies 
are slightly higher in this case. Therefore the
efficiency  values obtained with the matrix element
are used.

\end{itemize}

The overall relative systematic error on the
selection efficiencies ranges from 7\% to
16\% and from 7\% to 11\% for
stop and sbottom, respectively. 
This error 
and the uncertainty on the background normalisation,
dominated by MC statistics, as well as the quoted uncertainty
on two-photon background,
are incorporated~\cite{syst} in the final results.


\section{Results}
\Tabref{dm} summarises the number of selected data and
expected background events with different $\Delta M$
selections for all investigated channels.
A total of 35 and 18 candidates appear in the
$\qst \to \mathrm{c} \chna$ and  $\qsb \to \mathrm{b} \chna$ 
selections, whereas 33.1~$\pm$~4.3 and
13.5~$\pm$~3.3 are expected from the SM processes. The numbers of
$\qst \to \mathrm{b} \ell \snu$ and $\qst \to \mathrm{b} \tau \snu$
candidates are 9 and 18, compared with 11.3~$\pm$~3.0 and 21.4~$\pm$~4.4 
expected events.

The composition of the expected background into  two-fermion,
four-fermion 
and two-photon 
processes is given in \Tabref{bkgd}. 
When all the $\Delta M$ selections for all investigated channels
are applied, 59 events are retained. This is consistent
with 60.4~$\pm$~6.5 events expected from SM processes, mainly due to
two-photon interactions. Thus
no evidence for stop or sbottom is found and
upper limits are derived
on their production cross sections.

\begin{figure}[h]

\vspace*{-50mm}

\hspace*{15mm}  
\includegraphics[bb=33 33 548 471,width=120mm,height=100mm,
    clip=true,draft=false]{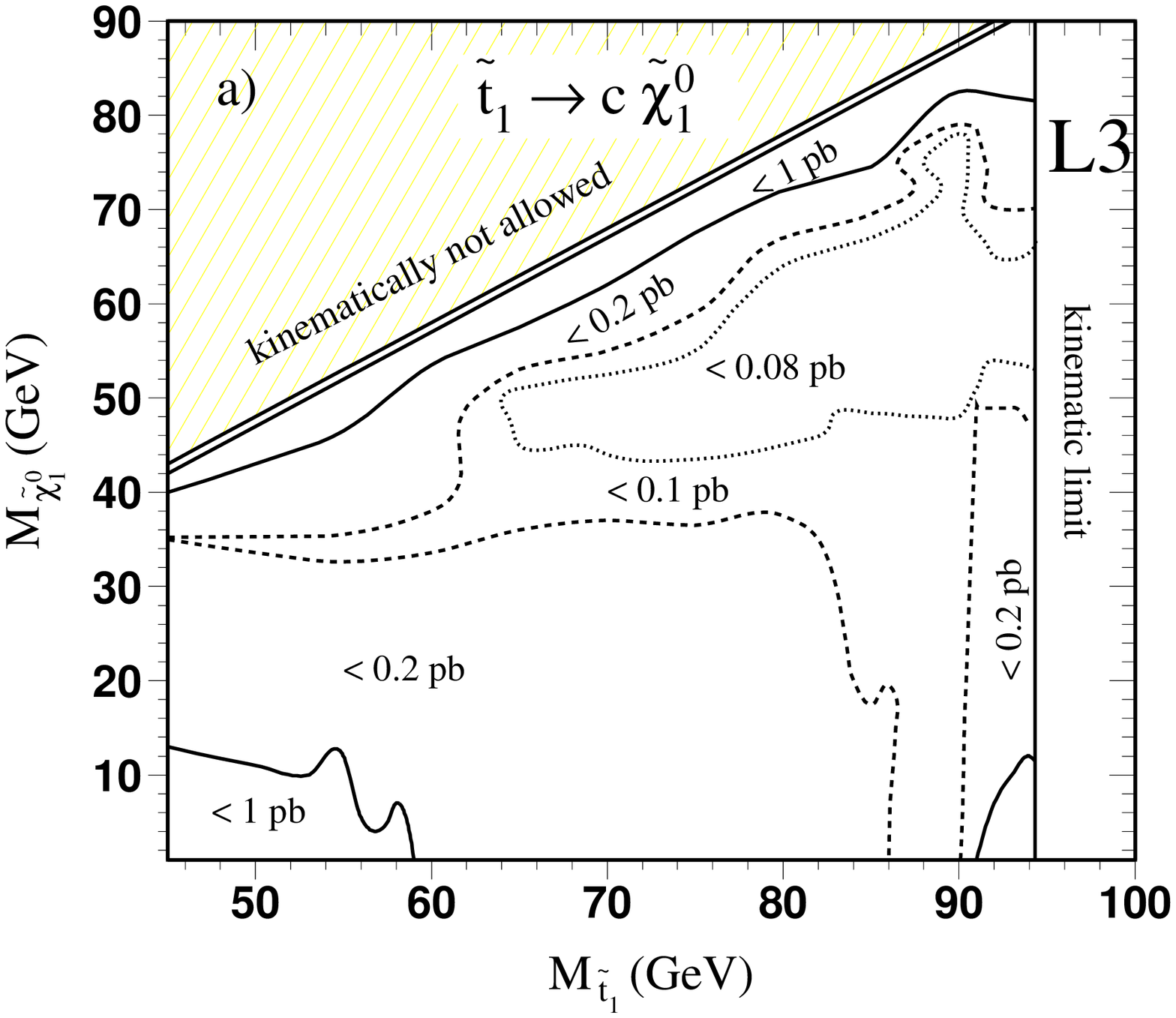}

\vspace*{10mm}

\hspace*{15mm}
\includegraphics[bb=33 33 548 471,width=120mm,height=100mm,
    clip=true,draft=false]{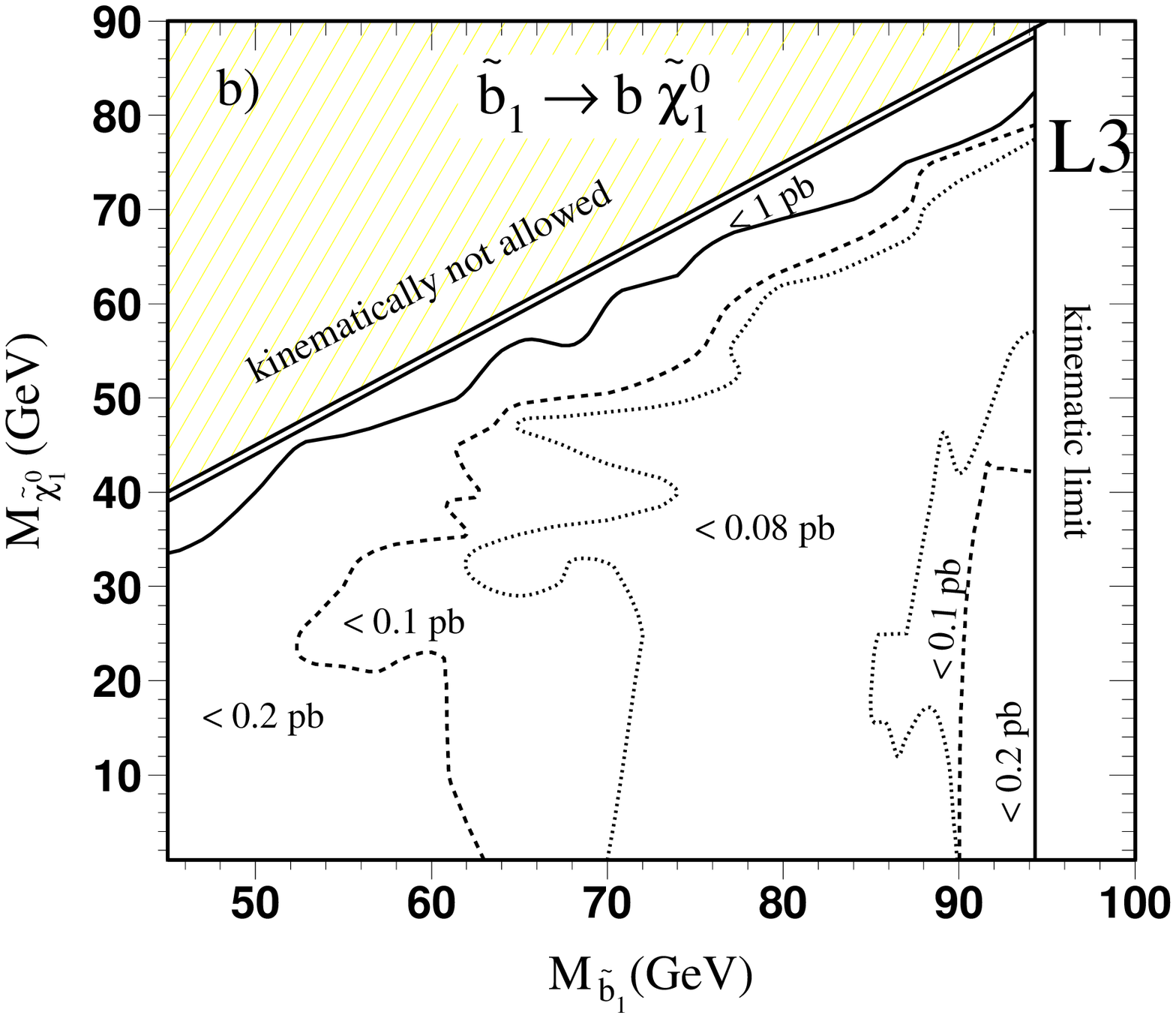}

\vspace*{5mm}

\caption{\label{fig:xbrexcl1}
Upper limits on a) 
$\mathrm{e^+e^-} \to \qst \qast \to \mathrm{c} \chna \mathrm{\bar{c}}  \chna$
and b) $\mathrm{e^+e^-} \to \qsb \qasb 
\to \mathrm{b} \chna \mathrm{\bar{b}} \chna$
production cross section times branching ratio. 
Limits are obtained by combining
the results at centre of mass energies of
$\sqrt{s}$=161--172 \gev{}, 183 \gev{} and 189 \gev{}.}
\end{figure}

Model-independent cross section limits
in the $M_{\mathrm{\tilde{q}}}$, $M_{\mathrm{\tilde{\chi}_1^0}}$ plane
are given in \Figref{xbrexcl1} for stop and sbottom 
assuming 100\% branching fraction
for the $\qst \to \mathrm{c} \chna$ and
$\qsb \to \mathrm{b} \chna$ decays.
The limits are obtained
by combining the present results with those
obtained at $\sqrt{s}=161-172$ \gev{} and 183 \gev{}
\cite{L3stop}.
The evaluated limits correspond to luminosity weighted
average cross sections. In the medium $\Delta M$ region
cross sections larger than 0.08 pb are excluded.

\begin{figure}[h]

\vspace*{-50mm}

\hspace*{15mm}
\includegraphics[bb=33 33 548 471,width=120mm,height=100mm,   
    clip=true,draft=false]{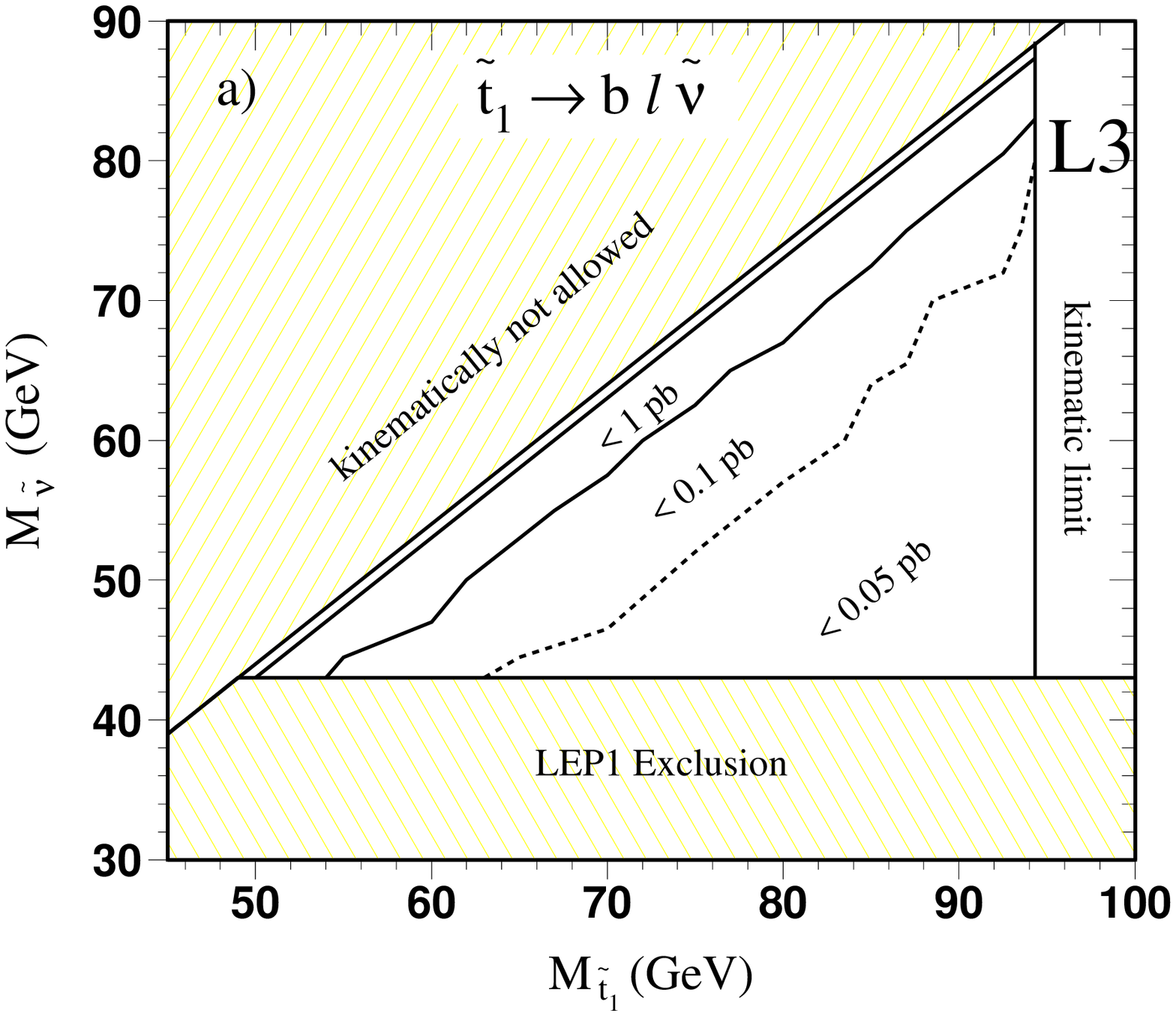}

\vspace*{10mm}

\hspace*{15mm}
\includegraphics[bb=33 33 548 471,width=120mm,height=100mm,
    clip=true,draft=false]{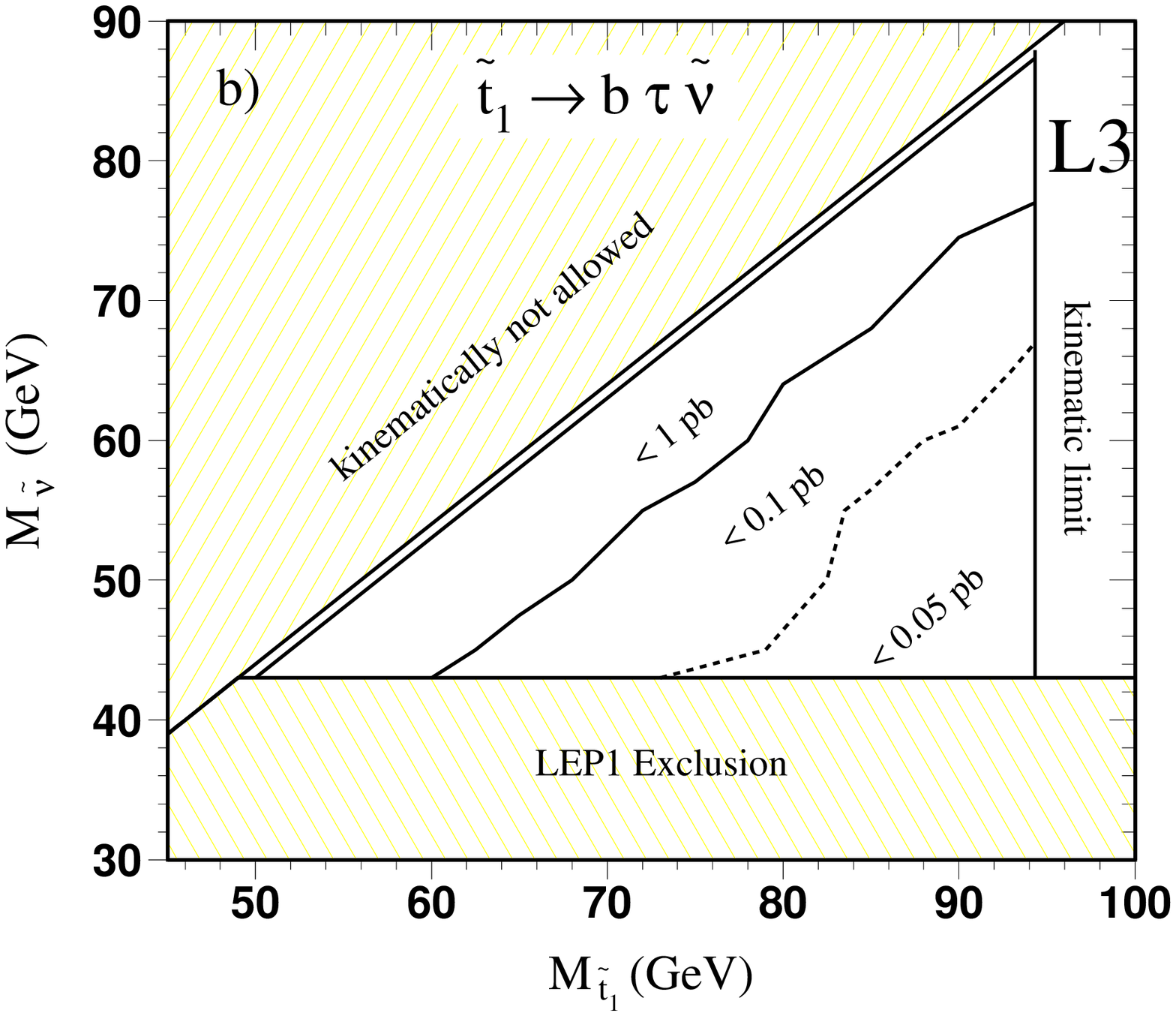}

\vspace*{5mm}

\caption{\label{fig:xbrexcl2}
Upper limits on a) $\mathrm{e^+e^-} \to
 \qst \qast \to 
\mathrm{b} \ell^+ \snu  \mathrm{\bar{b}} \ell^- \snu$,
$\ell=e, \mu, \tau$ assuming lepton universality
and b) 
 $\mathrm{e^+e^-} \to \qst \qast \to 
\mathrm{b} \tau^+ \snu \mathrm{\bar{b}} \tau^- \snu$
production cross section times branching ratio. Limits
are obtained from the $\sqrt{s}$=189 \gev{} data.}
\end{figure}

The cross section limits for stop production assuming 
$\qst \to \mathrm{b} \ell \snu$ decay, in the two scenarios for
lepton flavours, $\ell$=e, $\mu$, $\tau$ with equal probability
or $\ell$=$\tau$,
are given in \Figref{xbrexcl2}. Cross sections
larger than 0.05 pb are excluded if the mass difference
$\Delta M$ = $M_{\mathrm{\tilde{q}}} - M_{\snu}$
is greater than $25-35$ \gev{}.
%
\section{MSSM Interpretation}
\begin{figure}[h]

\vspace*{-60mm}

\hspace*{15mm}
\includegraphics[bb=33 33 548 471,width=120mm,height=95mm,
    clip=true,draft=false]{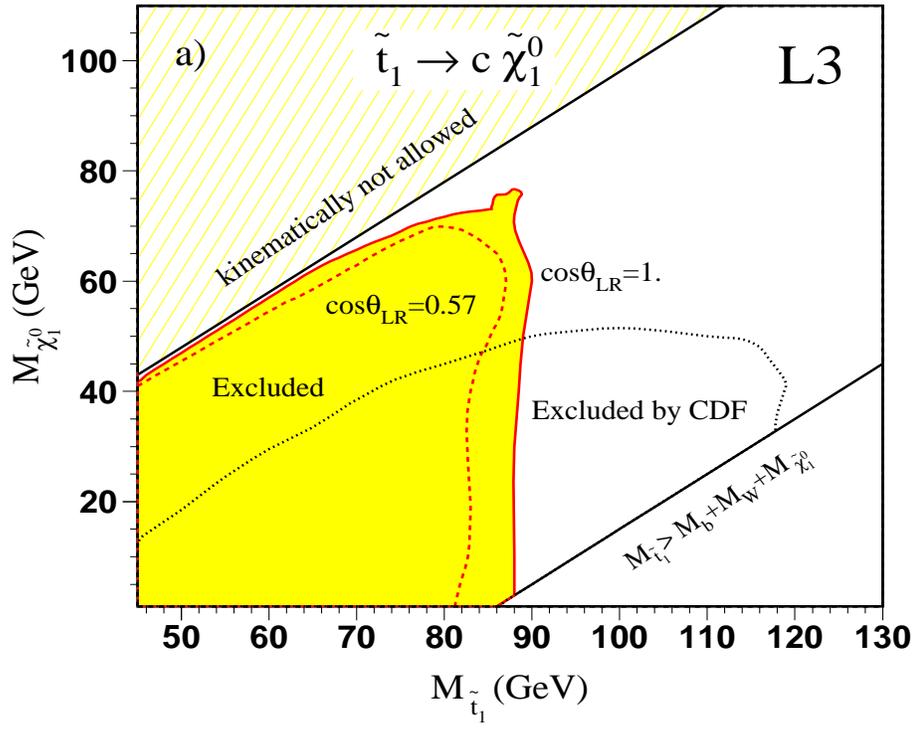}

\vspace*{10mm}

\hspace*{15mm}
\includegraphics[bb=33 33 548 471,width=120mm,height=95mm,
    clip=true,draft=false]{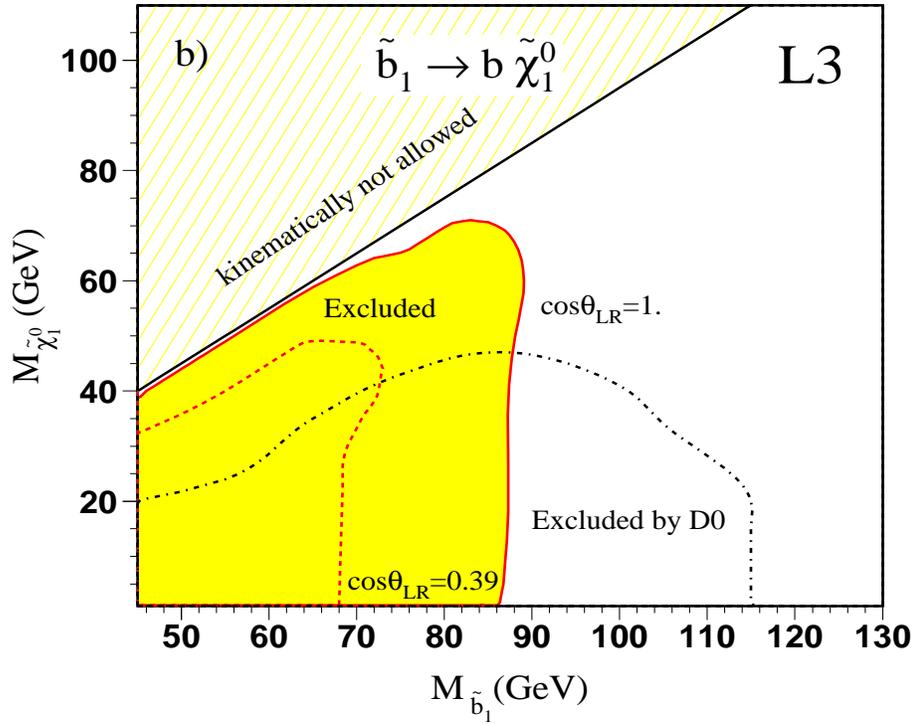}

\vspace*{5mm}

\caption{\label{fig:exclusion1}95\% C.L. exclusion 
limits in the MSSM on the masses of a) stop decaying via
$\qst \to \mathrm{c} \chna$
and b) sbottom decaying via
$\qsb \to \mathrm{b} \chna$
as a function of the
neutralino mass with maximal and minimal cross section
assumptions. 
For comparison results on stop 
searches obtained by CDF~\protect\cite{CDF} 
and on sbottom searches obtained by
D0~\protect\cite{D0} experiments are also shown.}
\end{figure}

In the MSSM the stop and sbottom production cross sections
depend on the squark mass  and the mixing angle
$\cos\theta_{\mathrm{LR}}$. 
Comparing the theoretical prediction 
with the 95\% C.L. limit on the production cross section,
we determine the excluded mass regions for $\qst$ 
and $\qsb$.
\Figref{exclusion1}a shows the excluded $\qst$   mass region as a
function of $M_{\qst}$ and $M_{\chna}$ 
at  $\cos\theta_{\mathrm{LR}}$=1 and 0.57 for the
$\qst \to \mathrm{c} \chna$ decay. For this decay mode,
stop masses below 88 \gev{} are excluded under the assumptions
of
$\Delta M$=$M_{\qst} - M_{\chna}$
greater than 15 \gev{} and $\cos\theta_{\mathrm{LR}}$=1. For the same
values of $\Delta M$ and in the most pessimistic scenario
of $\cos\theta_{\mathrm{LR}}$=0.57, the mass limit is 81 \gev{}.
The region where $\qst \to \mathrm{b} \mathrm{W} \chna$ 
decay is kinematically accessible and
becoming the dominant decay mode,
is also indicated. This decay is not considered in the analysis.

The exclusion plot for the sbottom is given in
\Figref{exclusion1}b for $\cos\theta_{\mathrm{LR}}$=1 and
$\cos\theta_{\mathrm{LR}}$=0.39.
Sbottom masses below 85 \gev{} are excluded assuming
$\Delta M$ greater than 15 \gev{} and $\cos\theta_{\mathrm{LR}}$=1. 
In the most pessimistic scenario of 
$\cos\theta_{\mathrm{LR}}$=0.39, the mass limit
obtained is 64 \gev{}.

\begin{figure}[h]

\vspace*{-50mm}

\hspace*{15mm}
\includegraphics[bb=33 33 548 471,width=120mm,height=95mm,
    clip=true,draft=false]{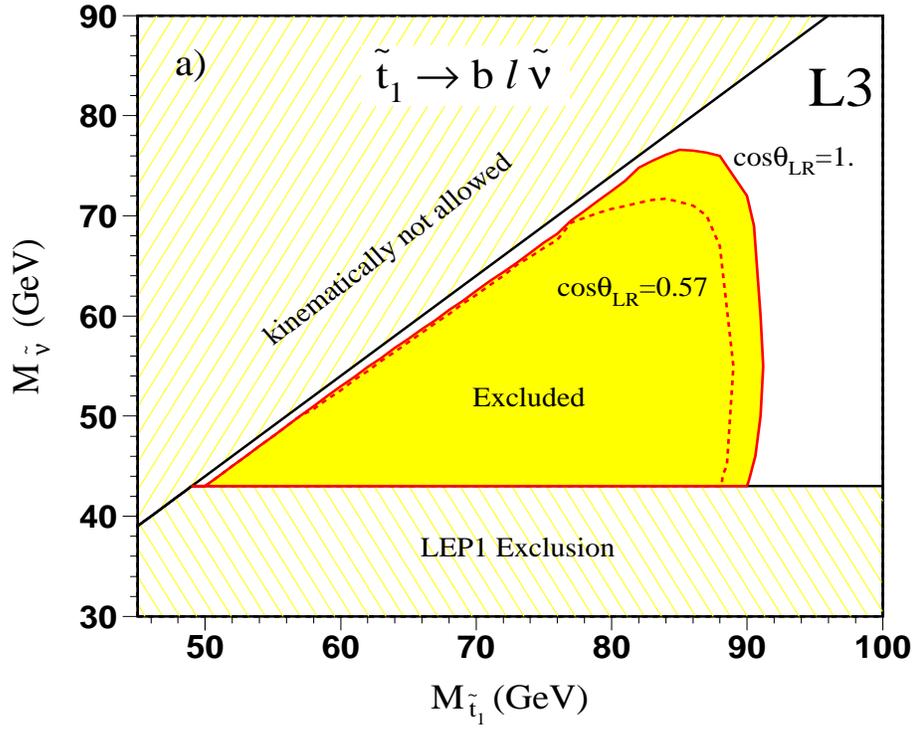}

\vspace*{10mm}

\hspace*{15mm}
\includegraphics[bb=33 33 548 471,width=120mm,height=95mm,
    clip=true,draft=false]{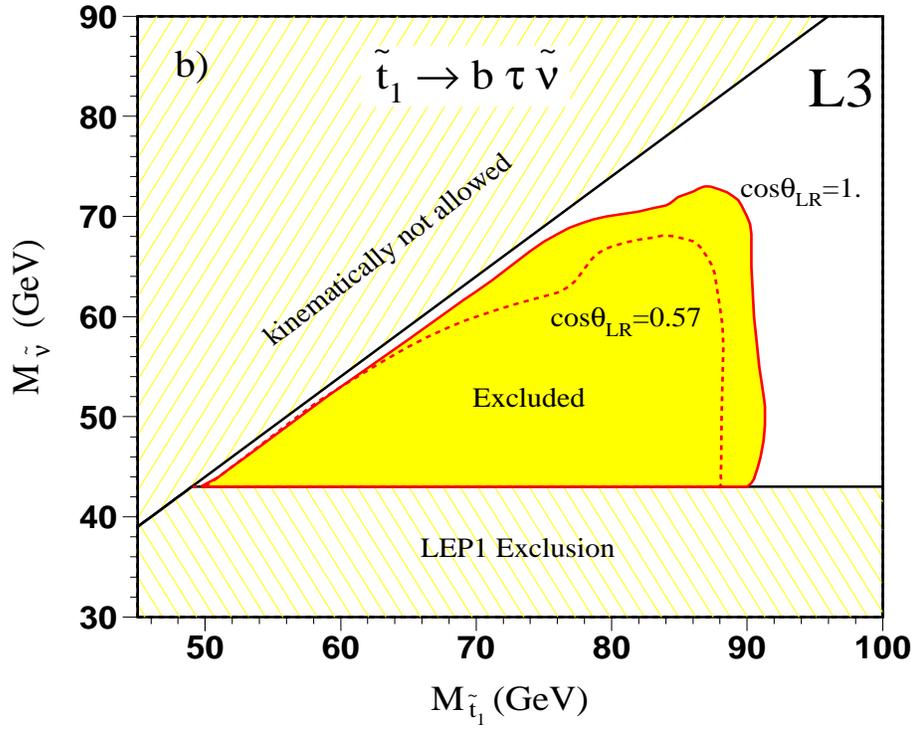}

\vspace*{5mm}

\caption{\label{fig:exclusion2}
95\% C.L. exclusion
limits in the MSSM on the mass of stop decaying via a)
$\qst \to \mathrm{b} \ell \snu$, $\ell=e, \mu, \tau$
with equal probability
and b) $\qst \to \mathrm{b} \tau \snu$,
as a function of the
sneutrino mass with maximal and minimal cross section
assumptions. The sneutrino mass limit
obtained at LEP1
is also shown.}
\end{figure}

The excluded stop mass regions, if the dominant
three-body decays are kinematically open,
are given in \Figref{exclusion2}.
\Figref{exclusion2}a corresponds to
$\qst \to \mathrm{b} \ell \snu$, 
$\ell$=e, $\mu$, $\tau$ with equal probability.
Here the lower $\qst$  mass limits are
89 \gev{} and 86 \gev{} for $\cos\theta_{\mathrm{LR}}$=1 and 0.57,
respectively.
The corresponding exclusion limits 
for stop decays through $\qst \to \mathrm{b} \tau \snu$
are shown in \Figref{exclusion2}b. 
Mass limits of 88 \gev{} and 83 \gev{} are obtained,
assuming $\Delta M >$15 \gev{}.

\begin{figure}[h]

\vspace*{-50mm}

\hspace*{15mm}
\includegraphics[bb=33 33 548 471,width=120mm,height=95mm,
    clip=true,draft=false]{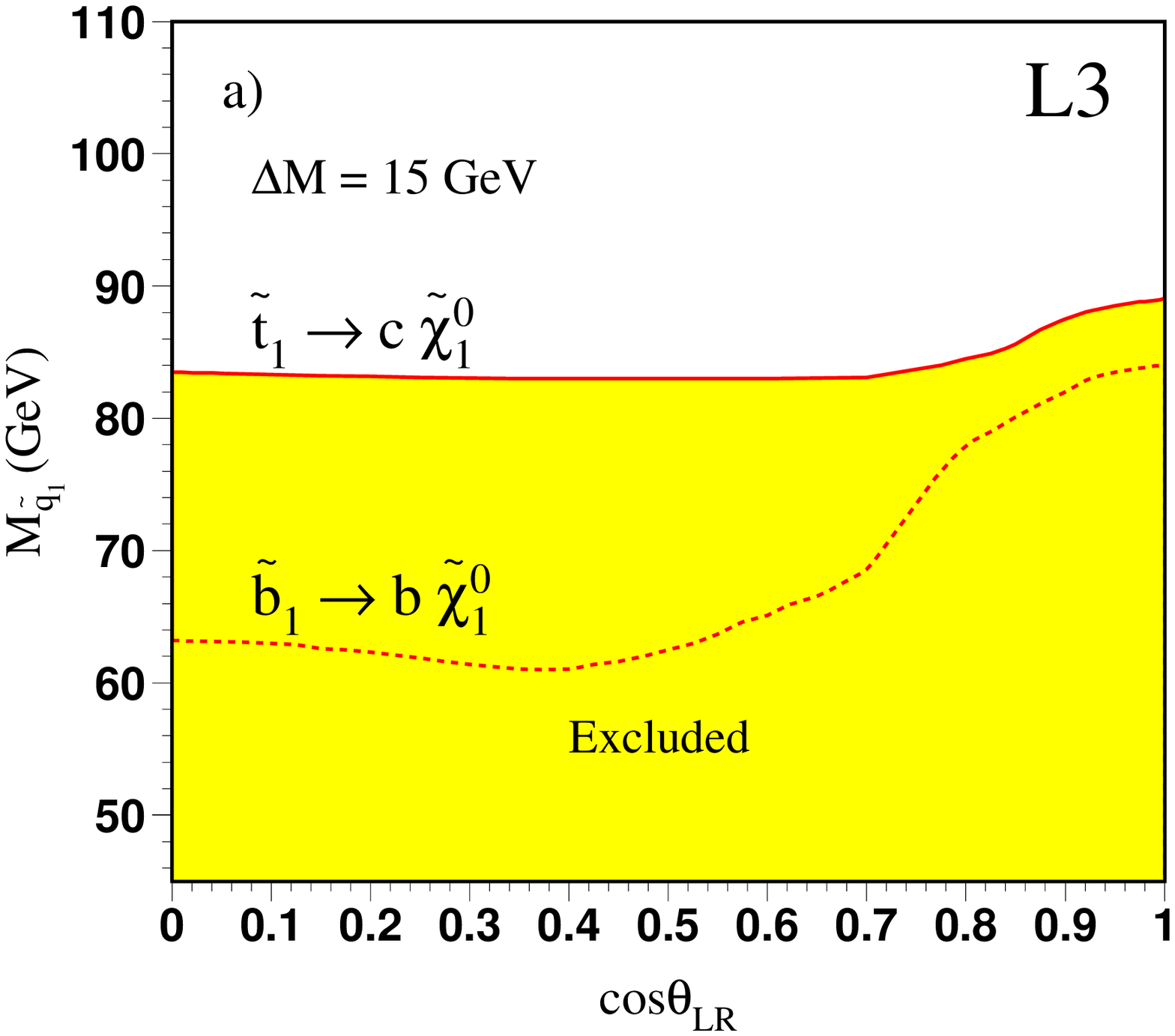}

\vspace*{10mm}

\hspace*{15mm}
\includegraphics[bb=33 33 548 471,width=120mm,height=95mm,
    clip=true,draft=false]{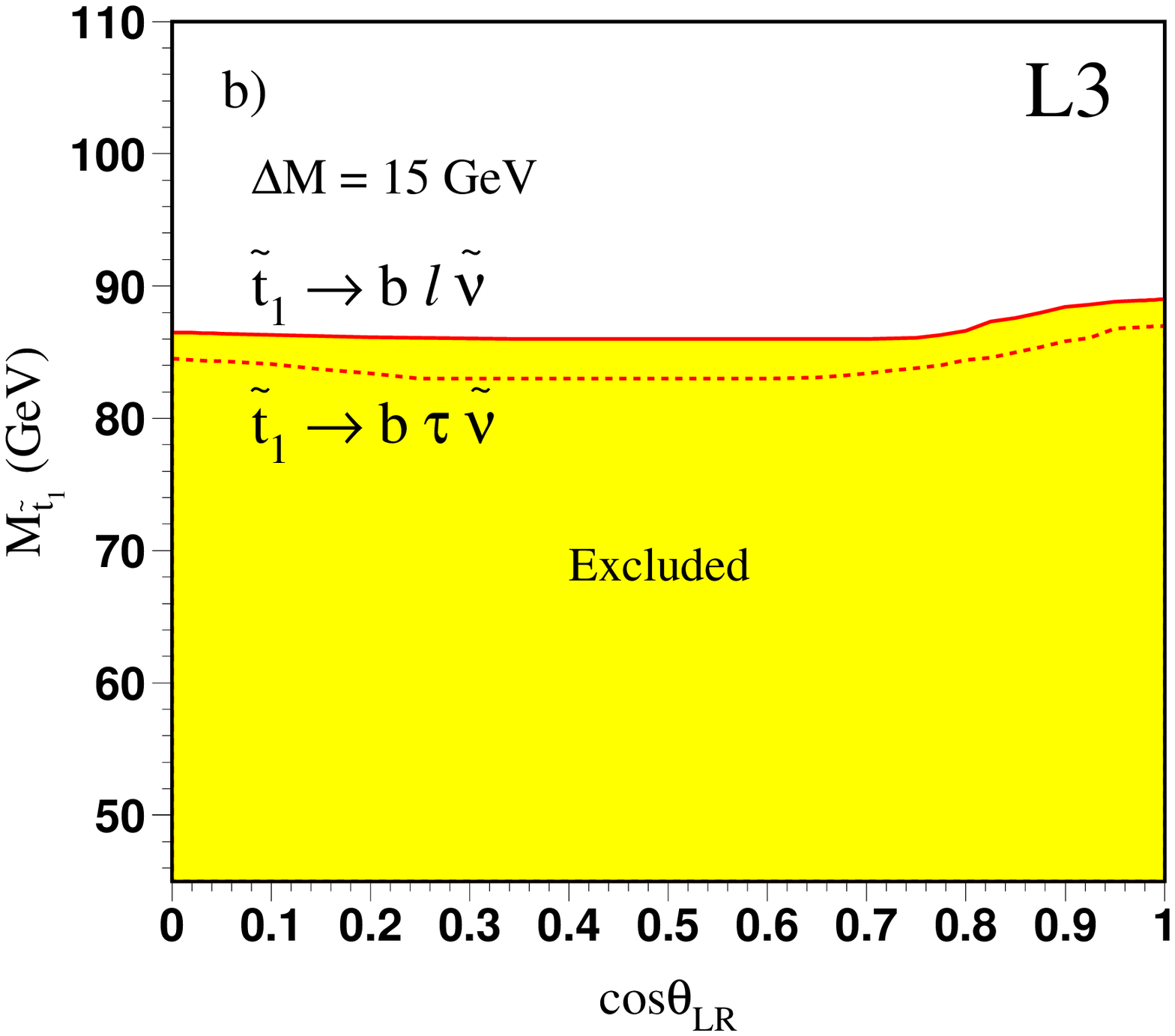}

\vspace*{5mm}

\caption{\label{fig:coslimits}  95\% C.L. exclusion limits 
in the MSSM
as a function of the mixing angle $\cos\theta_{\mathrm{LR}}$
for the 
a) stop decaying via $\qst \to \mathrm{c} \chna$ (solid line)
and sbottom decaying via $\qsb \to \mathrm{b} \chna$ (dashed
line),
b) stop decaying via $\qst \to \mathrm{b} \ell \snu$, $\ell=e, \mu,
\tau$
with equal probability (solid line)
and $\qst \to \mathrm{b} \tau \snu$ (dashed line).
}
\end{figure}

For a fixed value of $\Delta M$ =15 \gev{} the excluded stop 
and sbottom
masses as a function of the mixing angle
are shown in \Figref{coslimits}. 
The exclusion limits mainly
reflect the cross section behaviour.
At $\cos\theta_{\mathrm{LR}}$=1,
the $\qst$  and $\qsb$ cross sections are quite similar.
As $\cos\theta_{\mathrm{LR}}$ decreases squark production 
proceeds mainly via $\gamma$ exchange rendering
the sbottom production cross section about 4 times lower than that
of the stop. Consequently, the sbottom exclusion limits are
relatively
modest at low $\cos\theta_{\mathrm{LR}}$ values.

\begin{figure}[h]

\vspace*{-75mm}

\hspace*{15mm}
\includegraphics[bb=33 33 548 471,width=120mm,height=93mm,
    clip=true,draft=false]{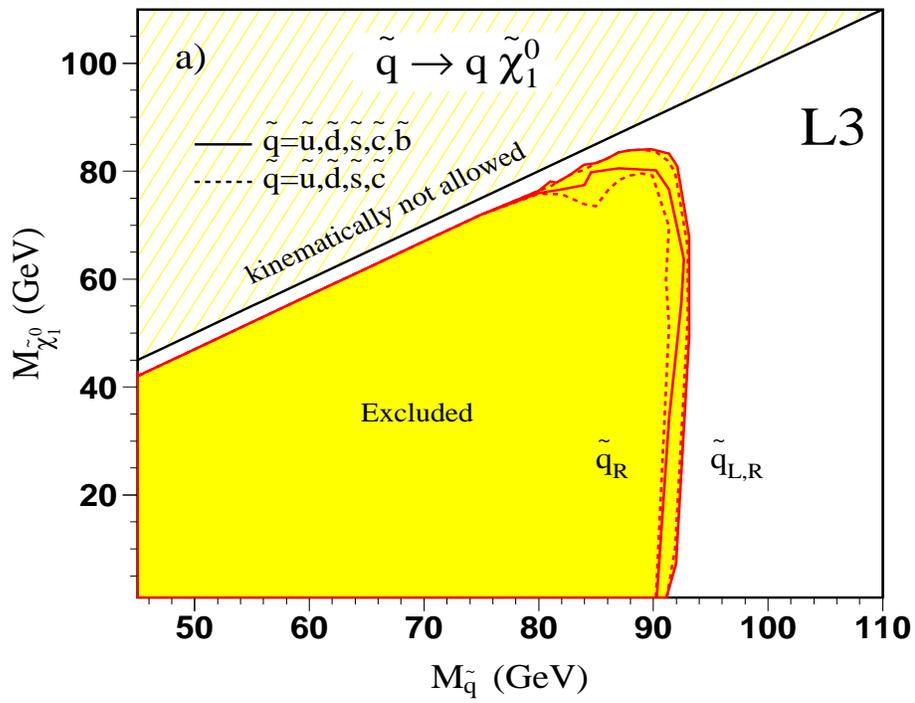}

\vspace*{10mm}

\hspace*{15mm}
\includegraphics[bb=33 33 548 471,width=120mm,height=93mm,
    clip=true,draft=false]{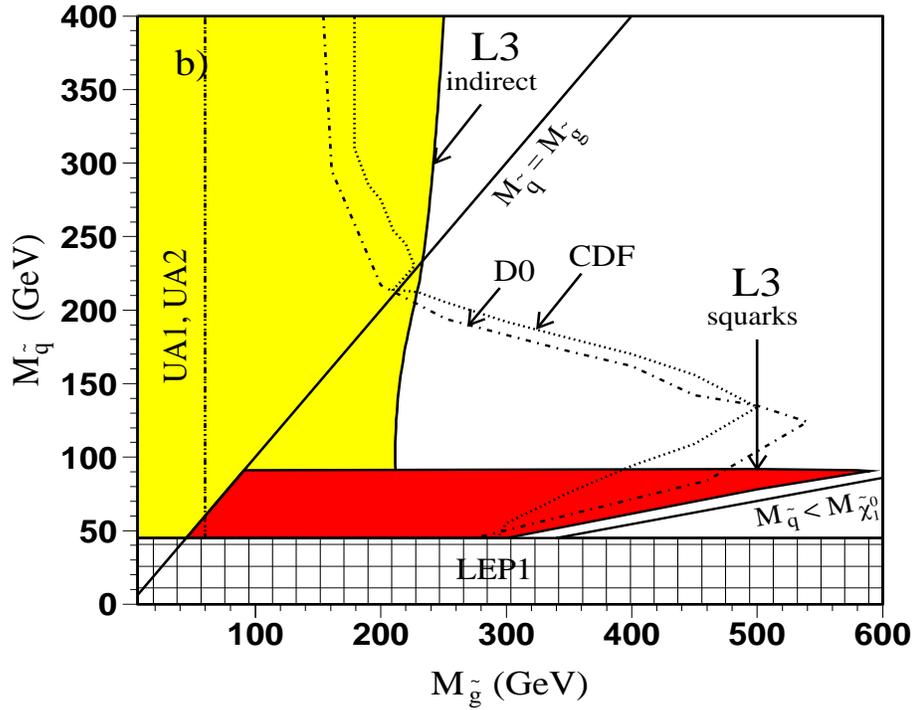}

\vspace*{5mm}

\caption{\label{fig:squarks}
a)  95\% C.L. exclusion limits in the MSSM on
the masses of the degenerate squarks decaying via 
$\tilde{\mathrm{q}} \to \mathrm{q} \chna$.
b) Excluded regions in the ($M_{\mathrm{\tilde{g}}}$,
$M_{\mathrm{\tilde{q}}}$)
plane. The dark shaded area is excluded from the
search of squarks of the first two families,
assuming the mass degeneracy among different flavours
and between ``left'' - ``right'' squarks.
The light shaded area illustrates
indirect limits on the gluino mass, derived from the
chargino, neutralino and scalar lepton searches.
The regions excluded by the CDF and D0 
collaborations~\protect\cite{CDFD0} are valid for
$\tan\beta=4$ and $\mu=-400$ \gev{}. The exclusions obtained by 
the UA1 and UA2~\protect\cite{ua1ua2}
collaborations are also shown.}
\end{figure}

For squarks of the first two generations,
the same selection efficiencies are assumed
as for the stop two-body
decays, because of the similar event topologies (jets and missing energy).
Then the cross section limits given in \Figref{xbrexcl1}a are
interpreted in terms of degenerate squark masses.
\Figref{squarks}a
shows the squark mass limit as a function of the
LSP mass. Two scenarios are considered:
``left'' and ``right'' squark  degeneracy  or 
only ``right'' squark production.
In the first case, with four degenerate squark flavours,
the mass limit is set at 91.5 \gev{} for  $\Delta M$
greater than 10 \gev{}. In the case of only
``right'' squark production, the  mass limit is
90 \gev{}.
The regions excluded, if all
squarks but the stop are degenerate are also shown.

Assuming gaugino unification at the GUT scale, the
results on the
four degenerate squarks are 
reinterpreted on the
$M_{\mathrm{\tilde{g}}}$, $M_{\mathrm{\tilde{q}}}$ plane as shown
in \Figref{squarks}b.
Moreover, the gaugino unification
allows a transformation of the absolute limit on
$M_2$,
obtained from the chargino, neutralino and scalar lepton
searches~\cite{chargino},
into a limit on the gluino mass as shown in \Figref{squarks}b.
This is done using the ISAJET program~\cite{isajet}.
For $\tan\beta=4$,
gluino masses up to about 210$-$250 \gev{} are excluded 
at 95\% C.L.

%
%
\section*{Acknowledgements}
We thank S.~Kraml for many useful discussions and comments.
We wish to express our gratitude to the CERN accelerator division for 
the excellent performance of the LEP machine. We acknowledge
the effort of those engineers and technicians who
have participated in the construction and maintenance of this
experiment.

%
%

\newpage
\bibliographystyle{l3stylem}

%
%

\newpage
\typeout{   }     
\typeout{Using author list for paper 186 -?}
\typeout{$Modified: Fri Sep 10 08:43:14 1999 by clare $}
\typeout{!!!!  This should only be used with document option a4p!!!!}
\typeout{   }
%
%
%
%
%
%

\newcount\tutecount  \tutecount=0
\def\tutenum#1{\global\advance\tutecount by 1 \xdef#1{\the\tutecount}}
\def\tute#1{$^{#1}$}
\tutenum\aachen            
\tutenum\nikhef            
\tutenum\mich              
\tutenum\lapp              
\tutenum\basel             
\tutenum\lsu               
\tutenum\beijing           
\tutenum\berlin            
\tutenum\bologna           
\tutenum\tata              
\tutenum\ne                
\tutenum\bucharest         
\tutenum\budapest          
\tutenum\mit               
\tutenum\debrecen          
\tutenum\florence          
\tutenum\cern              
\tutenum\wl                
\tutenum\geneva            
\tutenum\hefei             
\tutenum\seft              
\tutenum\lausanne          
\tutenum\lecce             
\tutenum\lyon              
\tutenum\madrid            
\tutenum\milan             
\tutenum\moscow            
\tutenum\naples            
\tutenum\cyprus            
\tutenum\nymegen           
\tutenum\caltech           
\tutenum\perugia           
\tutenum\cmu               
\tutenum\prince            
\tutenum\rome              
\tutenum\peters            
\tutenum\salerno           
\tutenum\ucsd              
\tutenum\santiago          
\tutenum\sofia             
\tutenum\korea             
\tutenum\alabama           
\tutenum\utrecht           
\tutenum\purdue            
\tutenum\psinst            
\tutenum\zeuthen           
\tutenum\eth               
\tutenum\hamburg           
\tutenum\taiwan            
\tutenum\tsinghua          
{
\parskip=0pt
\noindent
{\bf The L3 Collaboration:}
\ifx\selectfont\undefined
 \baselineskip=10.8pt
 \baselineskip\baselinestretch\baselineskip
 \normalbaselineskip\baselineskip
 \ixpt
\else
 \fontsize{9}{10.8pt}\selectfont
\fi
\medskip
\tolerance=10000
\hbadness=5000
\raggedright
\hsize=162truemm\hoffset=0mm
\def\r{\rlap,}
\noindent

M.Acciarri\r\tute\milan\
P.Achard\r\tute\geneva\ 
O.Adriani\r\tute{\florence}\ 
M.Aguilar-Benitez\r\tute\madrid\ 
J.Alcaraz\r\tute\madrid\ 
G.Alemanni\r\tute\lausanne\
J.Allaby\r\tute\cern\
A.Aloisio\r\tute\naples\ 
M.G.Alviggi\r\tute\naples\
G.Ambrosi\r\tute\geneva\
% ...
\newpage
\begin{list}{A}{\itemsep=0pt plus 0pt minus 0pt\parsep=0pt plus 0pt minus 0pt
                \topsep=0pt plus 0pt minus 0pt}
\item[\aachen]
 I. Physikalisches Institut, RWTH, D-52056 Aachen, FRG$^{\S}$\\
 III. Physikalisches Institut, RWTH, D-52056 Aachen, FRG$^{\S}$
\item[\nikhef] National Institute for High Energy Physics, NIKHEF, 
     and University of Amsterdam, NL-1009 DB Amsterdam, The Netherlands
\item[\mich] University of Michigan, Ann Arbor, MI 48109, USA
\item[\lapp] Laboratoire d'Annecy-le-Vieux de Physique des Particules, 
     LAPP,IN2P3-CNRS, BP 110, F-74941 Annecy-le-Vieux CEDEX, France
\item[\basel] Institute of Physics, University of Basel, CH-4056 Basel,
     Switzerland
\item[\lsu] Louisiana State University, Baton Rouge, LA 70803, USA
\item[\beijing] Institute of High Energy Physics, IHEP, 
  100039 Beijing, China$^{\triangle}$ 
\item[\berlin] Humboldt University, D-10099 Berlin, FRG$^{\S}$
\item[\bologna] University of Bologna and INFN-Sezione di Bologna, 
     I-40126 Bologna, Italy
\item[\tata] Tata Institute of Fundamental Research, Bombay 400 005, India
\item[\ne] Northeastern University, Boston, MA 02115, USA
\item[\bucharest] Institute of Atomic Physics and University of Bucharest,
     R-76900 Bucharest, Romania
\item[\budapest] Central Research Institute for Physics of the 
     Hungarian Academy of Sciences, H-1525 Budapest 114, Hungary$^{\ddag}$
\item[\mit] Massachusetts Institute of Technology, Cambridge, MA 02139, USA
\item[\debrecen] Lajos Kossuth University-ATOMKI, H-4010 Debrecen, Hungary$^\P$
\item[\florence] INFN Sezione di Firenze and University of Florence, 
     I-50125 Florence, Italy
\item[\cern] European Laboratory for Particle Physics, CERN, 
     CH-1211 Geneva 23, Switzerland
\item[\wl] World Laboratory, FBLJA  Project, CH-1211 Geneva 23, Switzerland
\item[\geneva] University of Geneva, CH-1211 Geneva 4, Switzerland
\item[\hefei] Chinese University of Science and Technology, USTC,
      Hefei, Anhui 230 029, China$^{\triangle}$
\item[\seft] SEFT, Research Institute for High Energy Physics, P.O. Box 9,
      SF-00014 Helsinki, Finland
\item[\lausanne] University of Lausanne, CH-1015 Lausanne, Switzerland
\item[\lecce] INFN-Sezione di Lecce and Universit\'a Degli Studi di Lecce,
     I-73100 Lecce, Italy
\item[\lyon] Institut de Physique Nucl\'eaire de Lyon, 
     IN2P3-CNRS,Universit\'e Claude Bernard, 
     F-69622 Villeurbanne, France
\item[\madrid] Centro de Investigaciones Energ{\'e}ticas, 
     Medioambientales y Tecnolog{\'\i}cas, CIEMAT, E-28040 Madrid,
     Spain${\flat}$ 
\item[\milan] INFN-Sezione di Milano, I-20133 Milan, Italy
\item[\moscow] Institute of Theoretical and Experimental Physics, ITEP, 
     Moscow, Russia
\item[\naples] INFN-Sezione di Napoli and University of Naples, 
     I-80125 Naples, Italy
\item[\cyprus] Department of Natural Sciences, University of Cyprus,
     Nicosia, Cyprus
\item[\nymegen] University of Nijmegen and NIKHEF, 
     NL-6525 ED Nijmegen, The Netherlands
\item[\caltech] California Institute of Technology, Pasadena, CA 91125, USA
\item[\perugia] INFN-Sezione di Perugia and Universit\'a Degli 
     Studi di Perugia, I-06100 Perugia, Italy   
\item[\cmu] Carnegie Mellon University, Pittsburgh, PA 15213, USA
\item[\prince] Princeton University, Princeton, NJ 08544, USA
\item[\rome] INFN-Sezione di Roma and University of Rome, ``La Sapienza",
     I-00185 Rome, Italy
\item[\peters] Nuclear Physics Institute, St. Petersburg, Russia
\item[\salerno] University and INFN, Salerno, I-84100 Salerno, Italy
\item[\ucsd] University of California, San Diego, CA 92093, USA
\item[\santiago] Dept. de Fisica de Particulas Elementales, Univ. de Santiago,
     E-15706 Santiago de Compostela, Spain
\item[\sofia] Bulgarian Academy of Sciences, Central Lab.~of 
     Mechatronics and Instrumentation, BU-1113 Sofia, Bulgaria
\item[\korea] Center for High Energy Physics, Adv.~Inst.~of Sciences
     and Technology, 305-701 Taejon,~Republic~of~{Korea}
\item[\alabama] University of Alabama, Tuscaloosa, AL 35486, USA
\item[\utrecht] Utrecht University and NIKHEF, NL-3584 CB Utrecht, 
     The Netherlands
\item[\purdue] Purdue University, West Lafayette, IN 47907, USA
\item[\psinst] Paul Scherrer Institut, PSI, CH-5232 Villigen, Switzerland
\item[\zeuthen] DESY, D-15738 Zeuthen, 
     FRG
\item[\eth] Eidgen\"ossische Technische Hochschule, ETH Z\"urich,
     CH-8093 Z\"urich, Switzerland
\item[\hamburg] University of Hamburg, D-22761 Hamburg, FRG
\item[\taiwan] National Central University, Chung-Li, Taiwan, China
\item[\tsinghua] Department of Physics, National Tsing Hua University,
      Taiwan, China
\item[\S]  Supported by the German Bundesministerium 
        f\"ur Bildung, Wissenschaft, Forschung und Technologie
\item[\ddag] Supported by the Hungarian OTKA fund under contract
numbers T019181, F023259 and T024011.
\item[\P] Also supported by the Hungarian OTKA fund under contract
  numbers T22238 and T026178.
\item[$\flat$] Supported also by the Comisi\'on Interministerial de Ciencia y 
        Tecnolog{\'\i}a.
\item[$\sharp$] Also supported by CONICET and Universidad Nacional de La Plata,
        CC 67, 1900 La Plata, Argentina.
\item[$\diamondsuit$] Also supported by Panjab University, Chandigarh-160014, 
        India.
\item[$\triangle$] Supported by the National Natural Science
  Foundation of China.
\item[\dag] Deceased.
\end{list}
}
\vfill






\newpage

\vspace*{40mm}
\begin{table}[h]
\begin{center}
\vspace*{-40mm}
\caption{\label{tab:effic} 
Selection efficiencies, $\epsilon$,
and number of expected events from SM processes,
$N_{\mathrm{SM}}$,  
for a 90 \GeV{} stop and sbottom, as a function of $\Delta M$ (see text).}
\vspace*{5mm}
\hspace*{-10mm}
\begin{tabular}{|c|c|c|c|c|c|c|c|c|}\hline
          & 
\multicolumn{2}{c|}{\ } & 
\multicolumn{2}{c|}{\ } & 
\multicolumn{2}{c|}{\ } & 
\multicolumn{2}{c|}{\ } \\*[-2mm]
$\Delta M$ (\GeV{})  & 
\multicolumn{2}{c|}{$\qst \to \mathrm{c} \chna$ } &
\multicolumn{2}{c|}{$\qst \to \mathrm{b}\ell \snu$ } &
\multicolumn{2}{c|}{$\qst \to \mathrm{b}\tau \snu$ } &
\multicolumn{2}{c|}{$\qsb \to \mathrm{b} \chna$ } 
\\*[2mm] \cline{2-9}
  & & & & & & & & \\*[-2mm]
  & $\epsilon$ (\%) & $N_{\mathrm{SM}}$  
  & $\epsilon$ (\%) & $N_{\mathrm{SM}}$ 
  & $\epsilon$ (\%) & $N_{\mathrm{SM}}$  
  & $\epsilon$ (\%) & $N_{\mathrm{SM}}$ 
\\*[2mm] \hline
 & &  & &  & &    & &      
\\*[-2mm]
2  &  
 0.1 & 17.7 &
  -  & -    & 
  -  & -    & 
  -  & - 
\\*[2mm]
5 &  
 17.5 & 17.7  & 
 -    & -     &
 -    & -     & 
0.06  & 12.3    
\\*[2mm]
7 &
21.6   &  21.8 &
15.8   &  10.7 &
5.6    & 12.3  &
17.6   & 12.3  
\\*[2mm]
10 &  
19.1   & 4.10 &
39.5   & 10.7 &
14.0   & 12.3 & 
14.5   & 12.7   
\\*[2mm]
20 & 
48.1   &  7.80  &
57.3   &  2.30  &
41.5   &  8.50  & 
35.4   &  0.46   
\\*[2mm]
30 &
62.7   & 4.37  &  
45.3   & 0.59  &    
35.2   & 1.58  &
42.8   & 0.73
\\*[2mm]
40 &
39.5   & 4.37  &  
46.0   & 0.59  &    
39.3   & 1.58  &
34.0   & 1.19  
\\*[2mm]
47 &
47.0   & 11.9  &
37.1   & 0.59  &
35.2   & 1.58  &
29.7   & 1.19
\\*[2mm]
60 & 
44.3  & 11.9  & 
-     & -     &   
-     & -     &
22.8  &  0.52 
\\*[2mm]
80 &
38.4 & 7.54  &
-    & -     &
-    & -     &
23.0 & 0.52   
\\*[2mm]
88 &
38.0  & 7.54 &
-     & -    &
-     & -    &   
21.6  &  0.52 
\\*[2mm]
\hline
\end{tabular}
\end{center}
\end{table}

\newpage

\vspace*{40mm}
\begin{table}[h]
\begin{center}
\vspace*{-40mm}
\caption{\label{tab:dm} Number of observed events,
$N_{\mathrm{D}}$,
and SM background expectations, $N_{\mathrm{SM}}$,
for the stop and sbottom selections at very low
(5--10 \gev{}), low (10--20 \gev{}), medium (20--40 \gev{}) and
high ($\gappeq$ 40 \gev{})
 $\Delta M$.
The quoted errors are due to MC statistics only.}
\vspace*{5mm}
\hspace*{-10mm}
\begin{tabular}{|r|c|c|c|c|c|c|c|c|}\hline
          & 
\multicolumn{2}{c|}{\ } & 
\multicolumn{2}{c|}{\ } & 
\multicolumn{2}{c|}{\ } & 
\multicolumn{2}{c|}{\ } \\*[-2mm]
Selection  & 
\multicolumn{2}{c|}{$\qst \to \mathrm{c} \chna$ } &
\multicolumn{2}{c|}{$\qst \to \mathrm{b}\ell \snu$ } &
\multicolumn{2}{c|}{$\qst \to \mathrm{b}\tau \snu$ } &
\multicolumn{2}{c|}{$\qsb \to \mathrm{b} \chna$ } 
\\*[2mm] \cline{2-9}
  & & & & & & & & \\*[-2mm]
  &  $N_{\mathrm{D}}$ & $N_{\mathrm{SM}}$  
  &  $N_{\mathrm{D}}$ & $N_{\mathrm{SM}}$ 
  &  $N_{\mathrm{D}}$ & $N_{\mathrm{SM}}$  
  &  $N_{\mathrm{D}}$ & $N_{\mathrm{SM}}$ 
\\*[2mm] \hline
 & &  & &  & &    & &   \\*[-2mm]
Very low   $\Delta M$  &  
 19  & 17.7~$\pm$~4.0 &
 7   & 8.4~$\pm$~2.7    & 
 14  & 12.3~$\pm$~3.4   & 
 16  & 12.3~$\pm$~3.3
\\*[2mm]
 & &  & &  & &    & &  \\*[-2mm]
Low   $\Delta M$ &  
 3  & 4.1~$\pm$~1.4  & 
 2  & 2.3~$\pm$~1.3    &
 4  & 8.5~$\pm$~2.7    & 
 0  & 0.46~$\pm$~0.22    
\\*[2mm]
 & &  & &  & &    & &  \\*[-2mm]
Medium   $\Delta M$ &
5   & 4.37~$\pm$~0.63 &
0   &  0.59~$\pm$~0.15 &
0    & 1.58~$\pm$~0.94  &
1   &  0.72~$\pm$~0.26  
\\*[2mm]
 & &  & &  & &    & &  \\*[-2mm]
High   $\Delta M$ &  
8   & 7.54~$\pm$~0.74 &
-   &  - &
-   &  - & 
2   & 0.52~$\pm$~0.14  
\\*[2mm]\hline
 & &  & &  & &    & &  \\*[-2mm]
Combined  &  
35   & 33.1~$\pm$~4.3 &
9   &  11.3~$\pm$~3.0 &
18   & 21.4~$\pm$~4.4 & 
18   & 13.5~$\pm$~3.3  
\\*[2mm]
\hline
\end{tabular}
\end{center}
\end{table}

\newpage

\vspace*{40mm}

\begin{table}[h]
\begin{center}
\vspace*{-40mm}
\caption{\label{tab:bkgd} Number of observed events,
$N_{\mathrm{D}}$,
and SM background expectations, $N_{\mathrm{SM}}$,
for the stop and sbottom selections.
The contribution of two-fermion ($\mathrm{q\bar{q}}$, $\tau^+ \tau^-$),
four-fermion ($\mathrm{W^+W^-}$,  $\mathrm{W^{\pm}e^{\mp}\nu}$, 
$\mathrm{ZZ}$,
$\mathrm{Ze^+e^-}$) and
two-photon ($\mathrm{e^+e^-q\bar{q}}$,  $\mathrm{e^+e^-\tau^+ \tau^-}$)
processes are given separately. The quoted errors are due to
MC statistics only.}
\vspace*{5mm}
\begin{tabular}{|l|c|c|c|c|c|}\hline
  & & & & &\\*[-2mm]
 Channel  & \ \ $N_{\mathrm{D}}$ \ \  &
$N_{\mathrm{two-fermion}}$ &
$N_{\mathrm{four-fermion}}$  &
$N_{\mathrm{two-photon}}$  
& \ \ $N_{\mathrm{SM}}$ \ \ 
\\*[2mm] \hline
  & & & & & \\*[-2mm]
$\qst \to \mathrm{c} \chna$  &  
35 & 
0.41~$\pm$~0.16  & 13.6~$\pm$~1.1 & 19.1~$\pm$~4.2 & 33.1~$\pm$~4.3
\\*[2mm] \hline
 & & & & & \\*[-2mm]
$\qst \to \mathrm{b}\ell \snu$ &  
9 & 
 0.29~$\pm$~0.15 & 0.97~$\pm$~0.24 & 10.0~$\pm$~3.0 & 11.3~$\pm$~3.0
\\*[2mm] \hline
 & & & & & \\*[-2mm]
$\qst \to \mathrm{b}\tau \snu$ &  
18 &
0.29~$\pm$~0.15 & 0.49~$\pm$~0.19 & 20.5~$\pm$~4.4 & 21.4~$\pm$~4.4
\\*[2mm] \hline
 & & & & & \\*[-2mm]
$\qsb \to \mathrm{b} \chna$ & 
18 &  
0.17~$\pm$~0.12 & 1.45~$\pm$~0.35 & 11.8~$\pm$~3.3 & 13.5~$\pm$~3.3 
\\*[2mm] \hline\hline
 & & & & & \\*[-2mm]
Total &
59 &
0.84~$\pm$~0.25 & 14.5~$\pm$~1.1 & 45.1~$\pm$~6.5 & 60.4~$\pm$~6.5
\\*[2mm] \hline
\end{tabular}
\end{center}
\end{table}

\newpage

\vspace*{40mm}

\newpage

\vspace*{40mm}

\newpage

\vspace*{40mm}   

\newpage

\vspace*{50mm}

\newpage

\vspace*{40mm}

\newpage

\vspace*{40mm}

\newpage

\vspace*{60mm}

\end{document}